\documentclass{article}

\pdfoutput=1
\usepackage{amsmath,amsfonts,amssymb,times,graphicx,natbib,algorithm,algorithmic,hyperref}

\usepackage[accepted]{data4good2016}

\newcommand{\myparagraph}[1]{\vspace{-.2cm}\paragraph{#1}}  

\newcommand{\squishlist}{
 \begin{list}{$\bullet$}
  { \setlength{\itemsep}{0pt}
     \setlength{\parsep}{3pt}
     \setlength{\topsep}{3pt}
     \setlength{\partopsep}{0pt}
     \setlength{\leftmargin}{1.5em}
     \setlength{\labelwidth}{1em}
     \setlength{\labelsep}{0.5em} } }

\newcommand{\squishlisttwo}{
 \begin{list}{$\bullet$}
  { \setlength{\itemsep}{0pt}
     \setlength{\parsep}{0pt}
    \setlength{\topsep}{0pt}
    \setlength{\partopsep}{0pt}
    \setlength{\leftmargin}{2em}
    \setlength{\labelwidth}{1.5em}
    \setlength{\labelsep}{0.5em} } }

\newcommand{\squishend}{
  \end{list}  }

\icmltitlerunning{Twitter as a Source of Global Mobility Patterns for Social Good}

\begin{document}

\twocolumn[
\icmltitle{Twitter as a Source of Global Mobility Patterns for Social Good}

\icmlauthor{Mark Dredze$^{*,\dagger}$}{mdredze@cs.jhu.edu}
\icmladdress{$^*$Human Language Technology Center of Excellence, Johns Hopkins University, Baltimore, MD 21211}
\icmlauthor{Manuel Garc\'{i}a-Herranz}{mgarciaherranz@unicef.org}
\icmlauthor{Alex Rutherford}{arutherford@unicef.org}
\icmladdress{UNICEF, 3 United Nations Plaza, New York, NY 10017}
\icmlauthor{Gideon Mann}{gmann16@bloomberg.net}
\icmladdress{$^{\dagger}$Bloomberg LP, 731 Lexington Ave, New York, NY 10022}

\icmlkeywords{data for good, travel statistics, Twitter}

\vskip 0.3in
]

\begin{abstract}
Data on human spatial distribution and movement is essential for understanding and analyzing social systems. However existing sources for this data are lacking in various ways; difficult to access, biased, have poor geographical or temporal resolution, or are significantly delayed.  In this paper, we describe how geolocation data from Twitter can be used to
estimate global mobility patterns and address these shortcomings. These findings will inform how this novel data source can be harnessed to address humanitarian and development efforts.
\end{abstract}


\section{Introduction}
\label{sec:introduction}
Social programs, whether developmental, humanitarian or public health related, rely on knowledge of where vulnerable populations are located. People travel nationally and internationally for a variety of purposes including regular commuting, seasonal work, tourism or coerced migration. Data describing these movements allow for the development of statistical models and analyses and significant universal patterns have been found in human mobility patterns \cite{Gonzalez:2008ul}.
Consider a few illustrative examples. In public health, epidemiological models of disease spread can forecast the course of an outbreak,allowing health workers to head-off infection transmission.
These models rely on travel data to project disease transmission between geographic
areas \cite{balcan2009multiscale,bajardi2011human,parker2011distributed,Viboud447,sadilek2012predicting}. Human migration, whether caused by economic hardship or resulting from physical danger to a population, follows both geographically advantageous routes, as well as previously established transit patterns resulting from existing connections between populations. Tracking current and predicting future migrations, which allows governments and NGOs to respond to migrations and refugee crises, depends on global mobility patterns \cite{JORS:JORS521,simini2012universal}.
A sudden change in established travel patterns may provide early-warning of crisis onset \cite{sonmez1998tourism,prideaux2003events}.

Researchers have utilized a diverse range of data resources for estimating global mobility patterns, each with distinct tradeoffs. These data sources can provide local travel patterns (within a metropolitan area) domestic patterns (travel within a country), and global patterns (international travel.) Airline travel dominates as a means for measuring
long distance travel \cite{colizza2006role,khan2009spread}. However, public air travel data is not timely, has poor coverage of local and domestic travel and may not accurately capture flights with connections\cite{iata}. Anonymized mobile phone meta-data can provide coverage of travel at multiple levels \cite{sagl2012social,krings2009urban,deville2014dynamic}
especially throughout urban areas \cite{calabrese2013understanding}.
As a result, mobile data has been used increasingly in epidemiological models of disease spread \cite{wesolowski2014commentary,bengtsson2015using}.
However, mobile data is proprietary and can be privacy sensitive. Diverse providers throughout
the world prevent the construction of a large, global dataset. Tourist statistics, available for many locations, do not typically
reflect traveler origin and tourism reflects only one type of travel. Finally, travel diaries, where people manually log
travel, are a traditional way of obtaining travel data \cite{axhausen1994travel}. But such methods simply cannot scale beyond specialized purposes motivating new approaches.

Social media provides a new and mostly untapped resource for obtaining travel patterns. Many social media platforms allow users to geotag their content. For example, Twitter allows users to geotag a tweet
with a specific set of coordinates -- using a GPS enabled device -- or tag a location as being associated with the message.
Additionally, local search and discovery services, such as Foursquare, allows users to {\em check-in} from different locations, creating
geotagged tweets.
The overall rate of geotagged tweets in Twitter remains low, roughly 2-3\%, but continues to grow. Even at this relatively low rate, with
roughly 500 million tweets per day, Twitter provides millions of geolocated data points on a daily basis.
Since most tweets are publicly available, the result is a large, public geotagged corpus.

There have been numerous uses of geotagged Twitter data, such as in public health \cite{Paul:2013lm,sadilek2012predicting}, political science \cite{o2010tweets}, linguistics \cite{eisenstein2010latent,eisenstein2014diffusion},
disaster response \cite{tapia2011seeking}, event detection \cite{watanabe2011jasmine}, topic discovery \cite{hong2012discovering} and location recommendation \cite{noulas2012random,liu2013point}. The importance of geotagged data has led to the task of geolocation,
in which a system automatically infers the location of a user
\cite{han2014text,rout2013s,compton2014geotagging,cha2015twitter,jurgens2015geolocation,osborne2014real,Dredze:2013a} or a specific tweet \cite{osborne2014real,Dredze:2016rm}.
Compared to the extensive literature on inferring and using geolocated Twitter data, there has been less work on understanding
aggregate location patterns. \cite{mocanu2013twitter} used location data to understand the languages of Twitter. \cite{leetaru2013mapping}  used geotagged tweets to describe the geography of Twitter. Some have studied check-in data, such as that from Foursquare, which provided
an early map of the emerging landscape of this type of data on Twitter \cite{cheng2011exploring,bauer2012talking}.
The most relevant work to ours is that of \cite{hawelka2014geo}, who also derived
global mobility patterns from Twitter. We contrast our work with theirs below.

This paper describes preliminary results from our investigation into Twitter as a data source of global mobility patterns for social good.
We consider a massive dataset: over 8.5 billion tweets that represent almost four years of all publicly available geotagged Twitter data.
We construct a global travel network for both cities and countries, which includes more than 87,856 cities and 248 countries, that reflects
travel patterns over four years.
We describe the construction of this travel network from Twitter data and a preliminary analysis of the resulting network.

\section{Data Resources}
\label{sec:data}
\myparagraph{Twitter} We use a collection of every publicly available geotagged tweet from January 1, 2012 to September 30, 2015. The collection contains 8.578 billion tweets from over 50 million users. Users had a median of 10 tweets each, with a mean of 168.98 and standard deviation of 962.5. Each tweet contains text, a time the tweet was posted, the user id and a location. These tweets include those authored directly by the user, or those created by an automated service, such as FourSquare. No private tweets were captured.

There are two methods by which users can share location information with their tweet. First, a user can author a message from a GPS enabled device, such as a smart phone. If geotagging is enabled, then the device will attach the current latitude and longitude to the tweet. Second, a user can choose to tag their tweet with a location. For example, a user may identify their location as ``Starbucks'' or ``Johns Hopkins University.'' In this case, Twitter associates a known location with the tweet. Locations can be countries, administrative areas (e.g. US States), cities, neighborhoods and points of interest (e.g. stores, parks, buildings, etc.). These locations contain several fields, including a name, location type and bounding box. Tweets may have both a set of coordinates and an associated known location. We note that users can also attach a location to their profile, which indicates their primary location, but we did not use this information on account of ambiguities e.g. `NYC/LA` or humorous locations `The World`.

\myparagraph{Geonames} While Twitter includes information about each location, we sought to map our data to an external knowledge resource. This will allow for future comparisons to other data sources, as well as inclusion of additional information about locations (e.g. populations, geographic administrative hierarchies, etc.) We use Geonames \cite{wick2012geonames}, a geographical database that covers all countries and contains over eight million named locations.
We used the datafile {\sc allCountries.zip}\footnote{Accessed April 25, 2016} which contains 11,005,123 locations. Each location comes with a set of coordinates and associated metadata (e.g. population).

\section{Computing Travel Statistics}
\label{sec:methods}
For each user in the Twitter collection, we organized all of their tweets from the entire time period chronologically. We then examined successive tweets to identify possible travel events as indicated by different locations between two adjacent tweets. A travel event is defined using the following guidelines:
\squishlisttwo
\item The successive tweets must occur within 72 hours of each other.
\item Both tweets must have a location as either a tagged location or specific coordinates.
\item The locations associated with each tweet are different, and one location does not contain the other. For example, a user may tweet first from Midtown Manhattan, and then from New York City. We would identify this as the same location.
\item The tweets must have been authored more than 50km apart. Tweets closer than 50km are not recognized as a travel event as they likely indicate local travel, even when they are in different locations. When specific coordinates are not available, distance is measured from the centroid of the bounding box of the associated location.
\squishend

Resulting travel events are associated with a timestamp (the time of the second tweet), a user, an origin and a destination location. Overall, we identified over 300 million travel events,
with the number of travel events per user having a median of 0, mean of 3.6 and standard deviation of 250.7.

\begin{figure}[!tb]

  \centering
  \includegraphics[width=.4\textwidth]{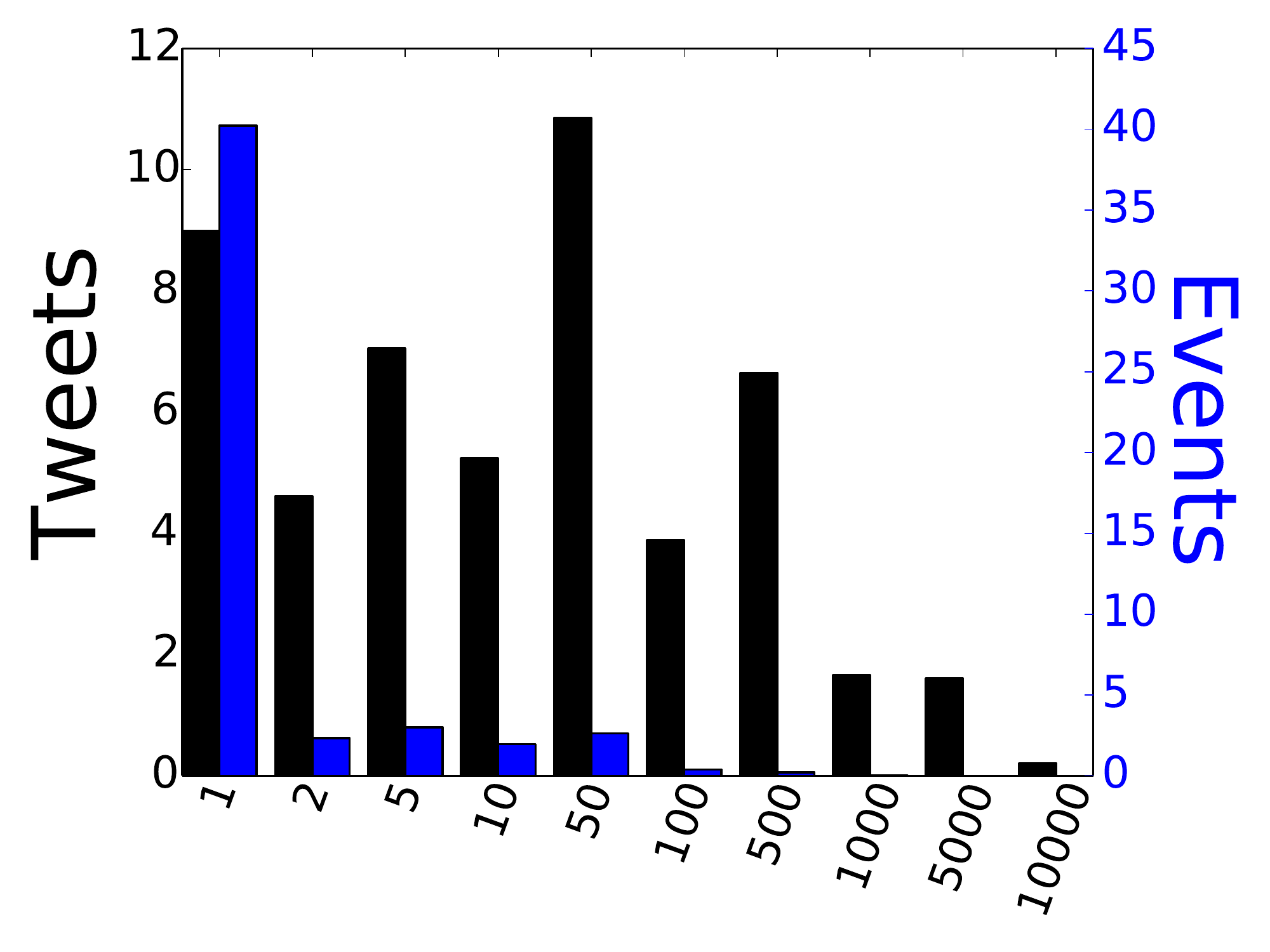}

\caption{\label{fig:tweet_dist} A histogram of the number of tweets (black) and events (blue) per user (y-axes in millions).}
\vspace{-.5cm}
\end{figure}

Some user accounts falsify location information for a variety of reasons. For example, a news aggregation account may list as its location the place most relevant to a tweeted story, or a spam account may attach false location information. We remove these accounts using several methods. First, we exclude travel events that require travel in excess of 1000km/hour (following \cite{hawelka2014geo,compton2014geotagging}). Second, as can be seen from the distribution of tweets per user (Figure (\ref{fig:tweet_dist})) there is a large skew in user activity. Therefore we remove users with more than 1000 geolocated tweets, roughly the top 4\% of all users in our data. Finally, we remove users who have more than 100 travel events, roughly the top 0.4\% of all users.

\myparagraph{Geonames Matching} We match every Twitter location to a Geonames location. We proceed in two passes. First, we attempt to match each location to a city with a population of at least 1,000 people (145,343 possible cities). Second, for unmatched places, we consider all possible locations in the database, which include administrative areas, roads, buildings, and other types of locations
(11,005,123 unique options). We match a Twitter location to a Geonames location by measuring the distance of the centroid of the Twitter
location, as computed from the provided bounding box, to the closest possible Geonames location, which is defined by a single set
of coordinates. We only consider matches closer than 50 km. Of the 1,128,662 unique Twitter locations 521 did not
match to Geonames; these locations were dropped from the data.

Table \ref{tab:geonames} shows statistics on the number and types of matches of Twitter locations to Geonames locations. We provide
statistics on the number of matches of unique Twitter location, as well as their coverage of the total dataset.

\begin{table}[t]
\begin{center}
\small
\begin{tabular}{|l|c|c|}
\hline
\bf Geonames Type & \bf Twitter Locations & \bf Travel Events \\
\hline
Admin area (A) & 672  & 115,163\\
Water (H) & 887 & 324,039 \\
Park (L) &  221 & 18,387 \\
City (P)&  84,982 & 156,273,217 \\
Road (R) & 12 & 33,012 \\
Point of interest (S) & 606 & 64,123 \\
Mountain (T) & 405 & 121,416 \\
Undersea (U) & 30 & 85,781 \\
Forest (V) & 4 & 339 \\
None & 26 & 352\\
\hline
\end{tabular}
\end{center}
\caption{ \label{tab:geonames} The types of Geonames locations used. Travel events include the type of each vertex on the edge in the count. Parenthesis indicate the Geonames feature type code.}
\vspace{-.5cm}
\end{table}
%

%
%
%
%
%

\myparagraph{Travel Network Construction}
The final step is to construct the travel network from the individual travel events.
We construct a graph, in which vertices are locations and weighted edges indicate the total number of travels between the two locations. We generate both a directed and undirected graph, where the undirected graph sums the weights of the two edges between a pair of vertices.

We construct two travel networks, where each has a directed and undirected version.
First, we use the Geonames locations to construct a full global network between cities and other types of Geonames locations. This network contains 7,688,854 edges between 87,856 vertices.
Second, we construct a global network between countries. This network contains 12,449 edges between 248 vertices. For this network, we rely on the country associated with each Twitter provided location, allowing for the inclusion of those few locations not successfully mapped to Geonames locations.
Statistics on each network are shown in Table \ref{tab:network_stats}.


\begin{table}[t]
\begin{center}
\small
\begin{tabular}{|l|c|c|}
\hline
\bf  & \bf Full Network & \bf Country Network \\
\hline
Locations & 87,856 & 248 \\
Edges & 7,688,854 & 12,449 \\
Edge density & 9.96$\times 10^{-4}$ & 1.6$\times 10^{-6}$ \\
Travel events & \multicolumn{2}{|c|}{309,253,447} \\
Total users &  \multicolumn{2}{|c|}{50,766,672} \\
Total tweets & \multicolumn{2}{|c|}{8,578,399,048} \\
\hline
\multicolumn{3}{|c|}{Only travel events} \\
\hline
Users  &\multicolumn{2}{|c|}{11,581,990}\\
Total tweets &  \multicolumn{2}{|c|}{2,145,810,039}\\
\hline
\end{tabular}
\end{center}
\caption{ \label{tab:network_stats} Statistics of the two travel networks.}
\vspace{-.3cm}
\end{table}

%
%
\begin{figure*}[!tb]
\begin{minipage}{0.49\textwidth}
  \centering
  \includegraphics[width=1\textwidth]{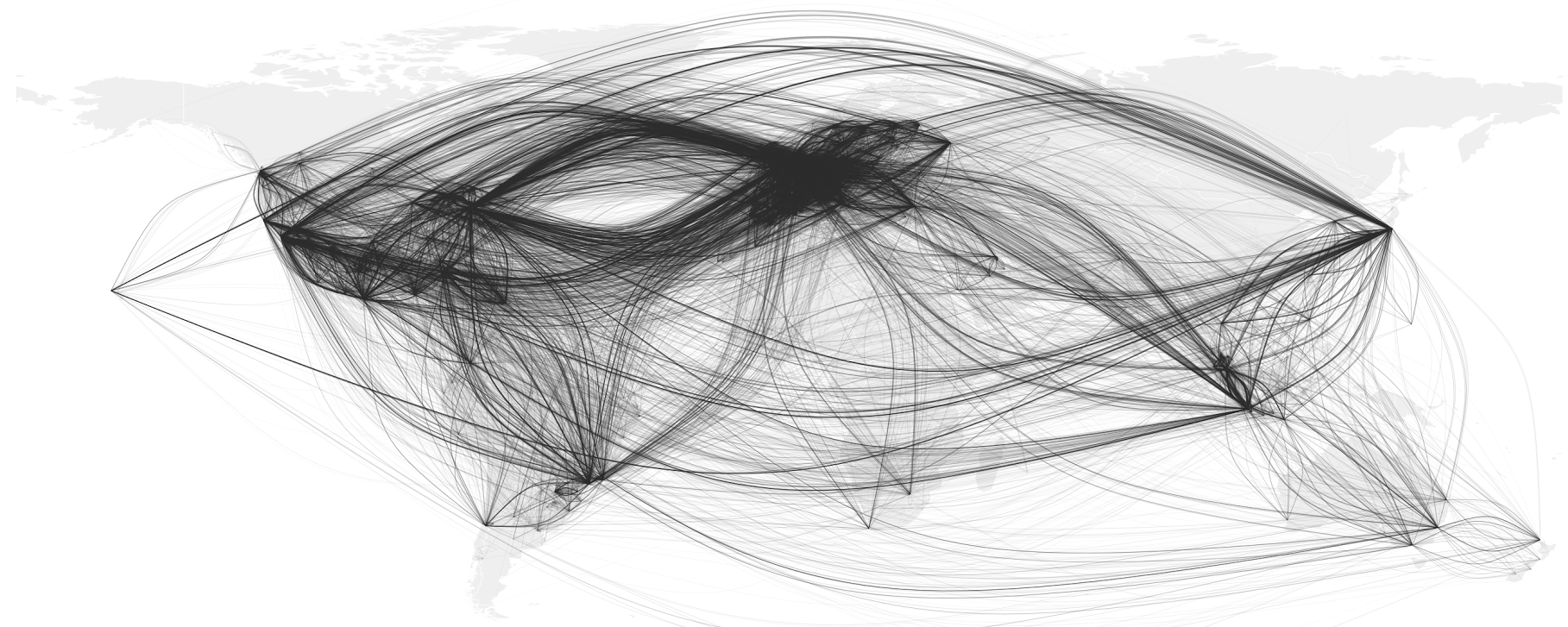}
\end{minipage}
\hspace{.3cm}
\begin{minipage}{0.49\textwidth}
  \centering
  \includegraphics[width=1\textwidth]{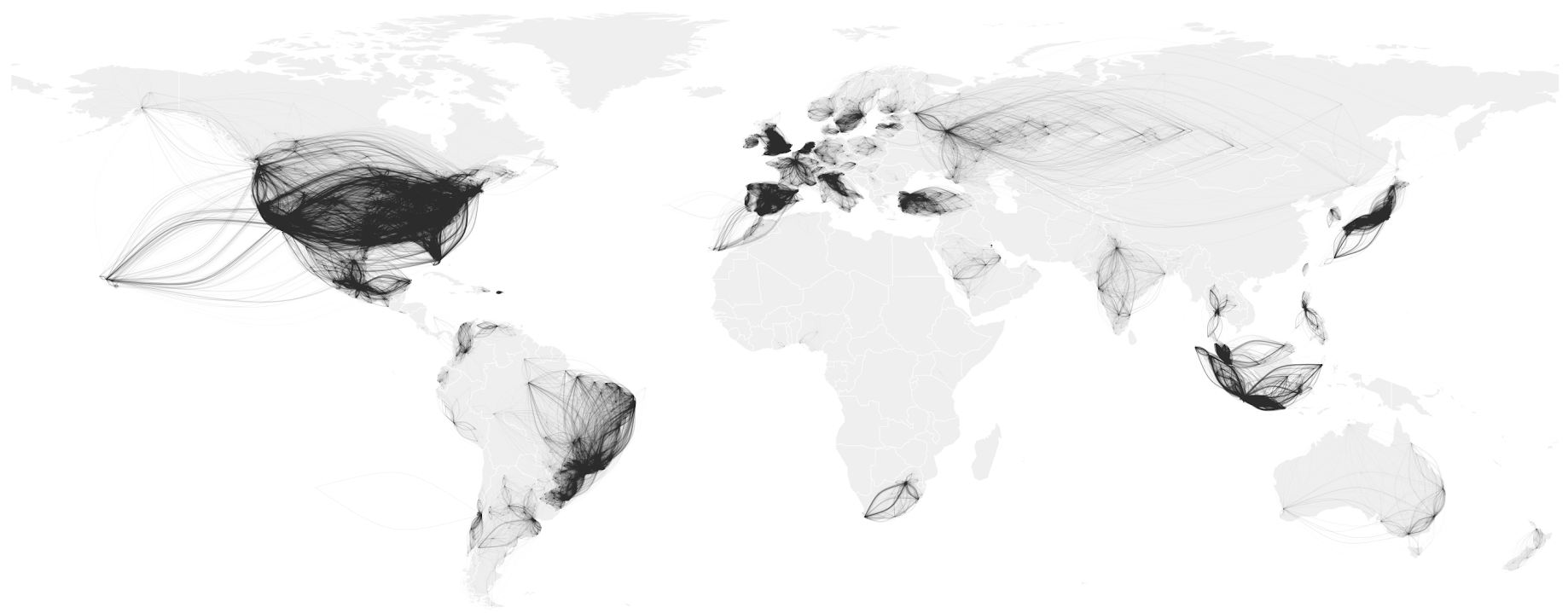}
\end{minipage}
\caption{\label{fig:maps} Mobility as observed in the travel network between countries (left) and between cities within the same country (right).}
\vspace{-.5cm}
\end{figure*}
%


\begin{table}[t]
\begin{center}
\scriptsize
\begin{tabular}{|c|c|c|c|}
\hline
\bf Continent &  \bf Top edge &\bf Top penetration (penetration)\\
\hline
Europe & UK-Spain & United Kingdom (4.4\%) \\
Africa & Botswana-Africa & South Africa (2.9\%) \\
North America & US-Canada & US (3.7\%) \\
South America & Argentina-Brazil & Chile (3.6\%) \\
Asia & Indonesia-Malaysia & Qatar (4.1\%) \\
\hline
\end{tabular}
\end{center}
\caption{ \label{tab:penetration} Most common edge and country with the highest Twitter penetration per continent (with at least 5000 users).}
\vspace{-.5cm}
\end{table}


\subsection{Comparison to Prior Work}
The work of \cite{hawelka2014geo} also derived global mobility patterns from Twitter. We follow their approach with some modifications,
such as mapping to an external reference (Geonames), the criteria for identifying travel events, and the spam removal method.
The major difference from our work is the amount of data considered. They use one year worth of geotagged tweets from 2012, which
encompasses 944 million tweets. In contrast, our dataset is roughly nine times larger, and covers four years worth of data.
The most immediate benefit of the increase in data size is our ability to consider cities, whereas their analysis only included countries.
Additionally, their work presents methods for normalizing data by Twitter penetration. Our development of normalization methods is ongoing
and the results in this paper do not yet reflect those efforts. Finally, they include extensive evaluations of their data, and we intend
to replicate several of their analyses.

\section{Analysis}

Figure \ref{fig:maps} shows the number of travel events between locations on a world map, where edge intensity denotes weight.
For clarity, we filter the edges to show just country links (left) and links of cities within the same country (right).
Next, we compute statistics on user penetration: the country with the highest Twitter penetration (number of Twitter users normalized by population) for each continent (Table \ref{tab:penetration}.)
We also include the most heavily travelled country to country link in each continent. While these initial results are promising, they highlight the need for careful normalization of the data based on Twitter
penetration (as in \cite{hawelka2014geo}.) Additionally, we may consider merging locations in the same metropolitan area
to smooth out local travel effects \cite{han2014text}.



\section{Discussion}
Our preliminary results suggest that Twitter may be a promising new data source for global mobility patterns and we plan to evaluate the suitability of this dataset for several applications. The most pressing consideration is the representativity of Twitter as determined by relatively low adoption of the service within low income countries. Careful calibration of movements aggregated from Twitter relative to the user base is required.

The benefits of Twitter as a mobility data source are clear. Firstly, data can be collected in real-time and are easily accessible through public APIs. Twitter can also capture movements on smaller spatial scales i.e. intra-urban, that are not captured by long distance travel records.

In this work we do not consider the content of tweets. While this content has been shown to be of great value in monitoring the opinions and topics of interest of vulnerable populations, further development of taxonomies and tools are required to analyze non-European languages and so provide insight in lower income countries. Analyzing these messages would allow us to
consider the relationship between topic and travel. For example, do users who discuss climate change take fewer long distance trips, or
are users who tweet about political activism less likely to travel to certain countries? We look forward to developing these ideas further
in future work.


{\bf Acknowledgements} We thank Adela Quinones and Mark Dimont for their insights and comments.

\begin{scriptsize}
\bibliography{refs_long}
\end{scriptsize}
\bibliographystyle{icml2016}

\end{document}